\documentstyle[12pt]{article}
\def\beq{\begin{equation}} 
\def\eeq{\end{equation}} 
\def\bea{\begin{eqnarray}}  
\def\eea{\end{eqnarray}}  
\def\bq{\begin{quote}}  
\def\eq{\end{quote}}  
\def\bi{\begin{itemize}}  
\def\ei{\end{itemize}}  
\def\beqa{\begin{eqnarray}}  
\def\eeqa{\end{eqnarray}}  
\def\be{\begin{enumerate}}  
\def\ee{\end{enumerate}}  
\def\beq{\begin{equation}}  
\def\eeq{\end{equation}}  
\def\bi{\begin{itemize}}
\def\ei{\end{itemize}}  
  
\def\grad{{\bf \nabla}}  
\def\pa{\partial}

\def\cp{{\cal P}}  
\def\cl{{\cal L}}

\def\r2{\sqrt{2}}  
\def\ra{\rightarrow}
\def\bi{\begin{itemize}}  
\def\ei{\end{itemize}}

\def\ov{\overline}  
\def\nn{\nonumber \\}

\def\ca{{\cal A}}

\begin{document}
\pagestyle{empty}
\begin{flushright}   CERN-TH/2000-366
\end{flushright}
\vskip 2cm
\begin{center}
{\huge  Four-Dimensional Supergravities\\
from Five-Dimensional Brane Worlds}
\vspace*{5mm} \vspace*{1cm} 
\end{center}
\vspace*{5mm} \noindent
\vskip 0.5cm
\centerline{\bf Adam Falkowski${}^1$, Zygmunt Lalak${}^{1,2}$
and Stefan Pokorski${}^{1}$}
\vskip 1cm
\centerline{\em ${}^{1}$Institute of Theoretical Physics}
\centerline{\em University of Warsaw, Poland}
\vskip 0.3cm
\centerline{\em ${}^{2}$Theory Division, CERN}
\centerline{\em CH-1211 Geneva 23}
\vskip2cm

\centerline{\bf Abstract}
\vskip .3cm
We give the explicit form of the  four-dimensional 
effective supergravity action, which describes low-energy physics 
of the Randall--Sundrum model with moduli fields 
in the bulk and charged chiral matter living on the branes.
The relation between 5d and 4d physics is explicit: the low-energy
 action is derived  from the compactification of a locally supersymmetric model in five dimensions. The presence of odd $Z_2$ parity scalars in the bulk 
gives rise to effective potential for the radion in four dimensions. 
We describe the mechanism of supersymmetry breaking mediation,
 which relies on non-trivial configuration of these  $Z_2$-odd  bulk fields.
Broken supersymmetry leads to stabilization of the interbrane distance.

\vskip1cm
\begin{flushleft}   
CERN-TH/2000-366\\
February  2001
\end{flushleft}
\newpage
\pagestyle{plain}

\section{Introduction}

There is a growing theoretical evidence that brane models with warped 
geometries 
may play a prominent role in understanding the variety of field theoretical 
incarnations of the hierarchy problem \cite{rs1,rs2,rs3}. 
Nonetheless, the initial hope that the mere presence of extra dimensions 
would be a natural tool to control mass scales in gauge theories 
coupled to gravity turned out to be 
premature. Various hierarchy 
issues have resisted numerous  
proposals, and revealed another, slightly disguised, face of the  
fine-tuning, well known from four dimensions (an example 
is the cosmological constant problem,
see \cite{flln} and references therein).
The most stumbling observation is that whenever one finds a flat 4d foliation 
as the solution of higher-dimensional Einstein equations, which seems to be 
necessary for the existence of a realistic 4d effective theory, 
it is accompanied by a special choice of various parameters in the 
higher-dimensional Lagrangian. The fine-tuning seems to be even worse in 5d than 
in 4d, since typically one must correlate parameters living on spatially 
separated branes. Then there appears immediately the problem of stabilizing 
these special relations against  quantum corrections. 

This situation has prompted the proposal \cite{bagger,gp,flp}, that it 
is 
a version of brane--bulk supersymmetry that may be able to explain 
apparent fine-tunings and stabilize hierarchies against quantum corrections.
And indeed, the brane--bulk supersymmetry turns out to correlate in the right 
way the brane tensions and bulk cosmological constant in the supersymmetric Randall--Sundrum model. 
In addition, supergravity is likely to be necessary to embed brane worlds 
in string theory. 
    
Hence, there are good reasons 
to believe  that supersymmetry is an important ingredient of the higher 
dimensional unification and the quest for consistent supersymmetric 
versions of
brane worlds goes on, see \cite{bagger,gp,flp,hiszp,flp2,kallosh,bc,bc2,kl,kl2,kb,
pmayr,duff,louis,zucker,bagger2,ls,clp,Nojiri:2001ji}. 
In earlier papers \cite{bc,bc2,kl,kl2,Maldacena:2000mw} the attention has been focused on models with 
solitonic (thick) branes and several no-go theorems were established (but 
see \cite{pmayr,duff,louis,clp}). 
Finally, in papers  \cite{bagger,flp,flp2,kallosh} explicit supersymmetric 
models with 
delta-type (thin) branes were constructed. 

The distinguishing feature of the pure supergravity Lagrangians 
proposed in \cite{flp} is imposing the $Z_2$ 
symmetry,
such that gravitino masses are $Z_2$-odd. An elegant formulation of the 
model is given in ref. \cite{kallosh}, where additional non-propagating 
fields are introduced to independently supersymmetrize the branes 
and the bulk. In the on-shell picture for these  fields the models 
of ref. \cite{kallosh} and refs. \cite{flp,flp2} are the same. 

In ref. \cite{kallosh} an extension of the model to include vector 
multiplets has been worked out. On the other hand, in refs. \cite{flp,flp2} 
it has 
been noted that supersymmetric Randall--Sundrum-type models can be generalized 
to include the universal hypermultiplet and gauge fields and matter on the 
branes. The 
Lagrangian of such a construction has been given in \cite{flp,flp2}.
This opens up a phenomenological avenue,
which we follow in the present paper, 
and allows us to study issues such as supersymmetry breaking and its transmission through the bulk. We want to stress that it is impossible to 
perform a trustworthy research of these issues without having a complete, 
explicit, locally supersymmetric model including matter on the branes 
embedded in extra dimensions. This is the main drawback of the 
phenomenological studies of supersymmetric brane worlds published so far. 
Further, it is our opinion that one should study thoroughly the classic, and 
in a sense minimal, version of the 5d brane worlds where charged matter and 
observable gauge interactions are confined to branes. The point is that 
putting matter and gauge fields into the 5d bulk amounts extending in a 
non-trivial 
way the very attractive MSSM.
Hence, for the time being we prefer to clarify the situation in the minimal
models with just neutral fields living in the bulk. 
On the other hand, populating the bulk with 5d gravitational fields alone
does not seem to be realistic. In typical string compactifications,
just on the basis of simple dimensional reduction of higher-dimensional 
supergravities, one expects neutral (with respect to the SM group) matter, namely moduli fields, 
to coexist in the 5d bulk with supergravity multiplet.
To represent such bulk matter, we choose to work in the present paper 
with bulk hypermultiplets, which couple also to the branes. In the explicit 
calculations which shall lead us to a consistent 4d supergravity model in 
the final section of this paper, we shall employ a universal hypermultiplet, 
which reduces to the dilatonic chiral multiplet in 4d.  

Still, it should be clear that putting gauge fields (and more) in the bulk 
is a viable alternative to our models, and as shown in \cite{ls}
it may also provide a mechanism to stabilize the extra dimensions. 

The final goal of this paper is to formulate the effective low-energy 
theory that describes properly  the physics of the warped 
five-dimensional models with  gauge sectors on the branes. 
On the way to four-dimensional theory we investigate supersymmetry, supersymmetry 
breakdown and moduli stabilization using five-dimensional tools. 
In particular, we show that in the class of models that we consider,
i.e. models without non-trivial gauge sectors in the bulk, 
unbroken $N=1$ local supersymmetry (classical solutions with four unbroken 
supercharges) implies vanishing of the effective cosmological 
constant.
We demonstrate  the link between vanishing of the 4d cosmological 
constant, minimization of effective potentials in 5d and 4d, and moduli 
stabilization in five-dimensional  supersymmetric models 
presented in this paper.     
We also discuss supersymmetry breaking due to a global obstruction
against the extension of bulk Killing spinors to the branes, which is a
phenomenon observed earlier in the Horava--Witten model in 11d and 5d.

First steps towards the 4d effective theory were made in \cite{bagger2},\cite{ls}
(where the K\"ahler function for the radion field was identified).
In the set-up considered in this paper, supersymmetry 
in 5d is first broken from eight down to four supercharges by the BPS 
vacuum wall, and then again broken spontaneously down to $N=0$ by a 
switching on of expectation values of sources living on the branes. 
The general strategy follows the one \cite{peskin-mirabelli,ovrut,elpp,elp} 
that led to the complete and accurate description of the low-energy 
supersymmetry breakdown in the Horava--Witten models, see
 \cite{peskin-mirabelli,elpp}. We are able to find  maximally symmetric solutions to the 5d equations of motion 
within our supersymmetric model, and deduce the K\"ahler potential, 
superpotential and gauge kinetic functions describing physics of corresponding vacua in four dimensions. It turns out that the warped background modifies 
in an interesting way the kinetic terms for matter fields and the gauge kinetic function on the warped wall. There also appears a potential for the radion superfield, its origin being a modulus-dependent prefactor multiplying 
the superpotential on the warped wall in the expression for the 
4d effective superpotential. We do not need to introduce any non-trivial 
gauge sector in the bulk to generate a potential for the $T$ modulus. It is interesting to note that the structure of the effective 4d supergravity is completely different from that of the no-scale models. 
In these the 4d cosmological constant vanishes, while $F_T$ is undetermined and sets the supersymmetry breaking scale. In our model $F_T$ vanishes, supersymmetry is broken by $F_S$ and non-zero cosmological constant is induced. 
The complete and phenomenologically relevant 4d $N=1$ supergravity 
model which we managed to construct in this paper, should finally 
facilitate 
a detailed investigation of the low-energy physics of 
warped compactifications.     

\section{Unbroken supersymmetry in\\ the brane-world scenarios}
We begin with a brief review of the original RS model. The action is that of   
 5d gravity on $M_4 \times S_1/Z_2$  with  negative cosmological constant :
\beq
\label{rsaction}
S=M^3 \int d^5x \sqrt{-g}\left (\frac{1}{2}R+6 k^2 \right )+
\int d^5x \sqrt{-g_{i}} \left (-\lambda_1 \delta(x^5)- \lambda_2\delta(x^5-\pi\rho) \right ). 
\eeq 
Three-branes  of non-zero tension are located at $Z_2$ fixed points.   The  ansatz for the vacuum solution preserving 4d Poincare invariance has the warped product form:
\beq 
\label{rsansatz}
ds^2= a^2(x^5)\eta_{\mu\nu}dx^\mu dx^\nu+R_0^2(dx^5)^2.
 \eeq
The breathing mode of the  fifth dimension is parametrized by $R_0$.
The solution for the warp factor $a(x^5)$ is: 
\beq
\label{exp}
a(x^5)=\exp (-R_0 k |x^5|).
\eeq
It  has an exponential form that can generate a large hierarchy of scales between the branes. Matching delta functions in the equations of motion requires fine tuning  of the brane tensions:
\beq
\label{ft}
\lambda_1=-\lambda_2=6k.
\eeq 
With the choice (\ref{ft}) the matching conditions are satisfied for arbitrary $R_0$, so the fifth dimension is not stabilized in the original RS model. Thus $R_0$ enters the 4d effective theory as a massless scalar (radion), which couples to gravity in manner of a Brans--Dicke scalar.  This is at odds with the precision tests of general relativity, so any realistic model should contain a potential for the radion field.    

Relaxing the condition (\ref{ft}) we are still able to find a solution in the maximally symmetric form, but only if we allow for  non-zero 4d curvature ($adS_4$ or $dS_4$) \cite{rk}. In such a case radion is stabilized and its vacuum expectation value is determined by the brane tensions and the bulk cosmological constant.  

The Randall-Sundrum model can be extended to a locally supersymmetric model \cite{bagger, gp, flp}. The basic set-up consists of 5d N=2  gauged supergravity \cite{gunaydin,agata} which includes the gravity multiplet $(e_\alpha^m, \psi_\alpha^A, \ca_\alpha)$, that is the metric (vielbein), a pair of symplectic Majorana gravitinos, and a vector field called the graviphoton.  The U(1) subgroup of the R-symmetry group is gauged, the gauge charge $g$ being  constant between two branes but  antisymmetric in the $x^5$ coordinate. The form of the gauging is fully characterized by the $SU(2)$ valued prepotential. Choosing the prepotential along the 
$\sigma^3$ direction:     
\beq
\label{su2prepotential}
g\cp=\frac{3}{2\r2} k \epsilon(x^5) i\sigma^3
\eeq
reproduces the bosonic part of the RS bulk action (\ref{rsaction}). Moreover, because of the antisymmetric function $\epsilon(x^5)$ , the supersymmetry variation of the action contains terms proportional to the delta function, that  cancel if the brane tensions satisfy the relation (\ref{ft}). Thus the fine-tuning present in the original RS model can be explained by the requirement of local  supersymmetry \cite{flp}.

 New bosonic and fermionic fields do not affect the vacuum solution
so the equations of motion  for the warp factor are the same as in the original, non-supersymmetric RS model. 
The RS solution  satisfies the BPS conditions and preserves one half of the supercharges, which corresponds to unbroken  N=1 supersymmetry in four dimensions.   
 In the supersymmetric version the brane tensions are fixed. As a consequence, the exponential solution (\ref{exp}) is the only maximally symmetric solution and the radion is still not stabilized.\\

We now turn to studying the supersymmetric RS model coupled to matter fields.  We want to investigate how general are the features present in the minimal supersymmetric RS model.  We find that unbroken local supersymmetry implies flat 4d space-time   in a wider class of 5d supergravities coupled to hyper- or vector multiplets, in which the scalar potential is generated by gauging a subgroup of R-symmetry.

Let us consider a version of the RS model, which apart from the gravity multiplet includes an arbitrary number of hypermultiplets.
 In five dimensions a hypermultiplet  consists of a pair of symplectic Majorana fermions $\lambda^a$ and of four real scalars $q^u$. The scalar potential can be generated by gauging a U(1) subgroup of the R-symmetry group  and at the same time, by gauging isometries of the sigma model \cite{gunaydin,ovrut,agata}. The bosonic action we consider is :
\bea
\label{hyperaction}
&S=M^3\int d^5x e_5\left(\frac{1}{2}R -h_{uv}D_\alpha q^u D^\alpha q^v - V(q)\right ) - \int d^4x e_4 \lambda_1 \delta(x^5)-\int d^4x e_4\lambda_2 \delta(x^5-\pi\rho)&
\\&
V(q)=g^2 (-\frac{16}{3}\vec{P}^2 + \frac{1}{2}h_{uv}k^u k^v)&
\eea 
The sigma-model metric $h_{uv}$ is quaternionic. The SU(2) valued prepotential $\cp=P_1i\sigma^1+P_2i\sigma^2+P_3i\sigma^3$ describes gauging of the R-symmetry group while the Killing spinor $k^u$ describes gauging the isometries of the quaternionic manifold; the covariant derivative acting on scalars is 
$D_\alpha q^u = \pa_\alpha q^u + gk^u\ca_\alpha$.
Generically, both $k^u$ and $\cp$ are functions of the hypermultiplet scalars and satisfy the 'Killing prepotential equation' $k^u K_{uw} = \pa_{w} \cp + [\omega_w,\cp]$, where $\omega$ is the spin connection and $K$ is the K\"ahler form of the quaternionic manifold . As usually, we assume that the gauge charge $g$ is odd: $g \equiv \frac{1}{\sqrt{2}}6k\epsilon(x^5)$. The brane tensions are:
\beq
\label{hyptensions}
 \lambda_1=-\lambda_2=24k P_3
\eeq 
as was pointed out in \cite{flp2}. This relation is  necessary to cancel the variation of the action which arise because of the presence of  $\epsilon(x^5)$ in the gauge coupling and is an equivalent of (\ref{ft}) in a model  with hypermultiplets. 

 The relevant part of the supersymmetry transformation laws is (we use the normalization of \cite{ovrut}):
\bea&
\label{susyhyp}
\delta \psi_\alpha^A =\pa_\alpha \epsilon^A+
 \frac{1}{4} \omega_{\alpha ab}\gamma^{ab} \epsilon^A+  
\frac{\r2}{3}\gamma_\alpha g P_3(\sigma^3)^A_B\epsilon^B   
&\nn
&\delta \lambda^a= -i V_u^{Aa}\pa_5q^u\gamma^5 \epsilon_A + g\frac{1}{\r2}k^u V_u^{Aa}\epsilon_A   
&\eea

Our objective is to show that, in the class of warped compactification,  unbroken supersymmetry implies flat 4d space-time. To achieve this,  we will derive the effective theory of the 4d metric degrees of freedom. It will turn out that supersymmetry requires vanishing of the 4d effective potential. We make the ansatz: 
\bea
\label{warpedansatz}&
ds^2=a^2(x^5)\bar{g}_{\mu\nu}dx^\mu dx^\nu+R_0^2 (dx^5)^2
&\nn&
q^u=q^u(x^5)
\eea
 which describes oscillations of the 4d metric $\bar{g}_{\mu\nu}$ about some vacuum solution.
Using the ansatz (\ref{warpedansatz}) the action (\ref{hyperaction}) can be rewritten in the form which reveals the BPS structure:
\bea
\label{bpsaction}
&S=M^3 \int d^5x\sqrt{-\bar{g}} R_0 \left( a^4 (
\frac{1}{2a^2}\bar{R} + \frac{6}{R_0^2}(\frac{a'}{a}+ 4kR_0\epsilon(x^5) \sqrt{\vec{P}^2})^2
+9k^2h_{uv}(4\pa^u \sqrt{\vec{P}^2} \pa^v \sqrt{\vec{P}^2}
-k^u k^v)
\right.&\nn &
-\frac{1}{R_0^2}h_{uv}(\pa_5 \phi^u -6 k R_0\epsilon(x^5) \pa^u \sqrt{\vec{P}^2})(\pa_5 \phi^v -6kR_0\epsilon(x^5) \pa^v \sqrt{\vec{P}^2}) 
-4\pa_5 (a^4 (\frac{a'}{a}+3k\epsilon(x^5)\sqrt{\vec{P}^2}))
&\nn & \left.
+a^4 (24k\sqrt{\vec{P}^2}-\lambda^1)\delta(x^5)+a^4 (-24k\sqrt{\vec{P}^2}-\lambda^2)\delta(x^5-\pi\rho)
\right)
&\eea 

To show that the effective potential vanishes when supersymmetry is preserved requires some calculations. First, $\delta \psi_\mu^A=0$ conditions can be solved in terms of the warp factor  yielding:
\bea 
&\frac{a'}{a}=-4kR_0 \epsilon(x^5) \sqrt{\vec{P}^2}
&\eea 
Furthermore, the $\delta\lambda^a=0$ condition yields:
\bea
\label{dlambdasolution}
&
V_u^{1a}\pa_5 q^u = i\frac{12\r2 kR_0}{4\sqrt{\vec{P}^2}}k^u(V_u^{1a}P_3+V_u^{2a}(P_1-iP_2))
&\nn&
V_u^{2a}\pa_5 q^u = i\frac{12\r2 kR_0}{4\sqrt{\vec{P}^2}}k^u(-V_u^{2a}P_3+V_u^{2a}(P_1+iP_2))
\eea
Using the Killing prepotential equation it is possible to eliminate $k^u$ from (\ref{dlambdasolution}). After some calculations we get:
\beq
h_{uw}\pa_5 q^w =6k R_0 \pa_u \sqrt{\vec{P}^2}
\eeq
Another  manipulations of (\ref{dlambdasolution}) yield the relation $h_{uv}\pa_5 q^u\pa_5 q^v= 9R_0^2 k^2 h_{uv} k^u k^v$ .

Summarizing, BPS conditions imply the following relations between the scalars, the warp factor and  the Killing vector and prepotential:
\bea
\label{hhBPS}
&
\frac{a'}{a}=-4 k R_0\sqrt{\vec{P}^2}
&\nn&
h_{uw}\pa_5 q^w = 6 k R_0 \pa_u \sqrt{\vec{P}^2} 
&\nn&
h_{uv}\pa_5 q^u\pa_5 q^v= 9 k^2 R_0^2 h_{uv} k^u k^v.
&\eea 
Plugging the above formulae into the action (\ref{bpsaction}) and integrating over $x^5$, we obtain the 4d effective action for the metric $\bar{g}_{\mu\nu}$:
\bea &
S_4=M_{PL}^2\int d^4x (\frac{1}{2}\bar{R}-V_{eff})
&\nn&
 M_{PL}^2 V_{eff}=- M^3 \left(a^4(0)(24 k \sqrt{\vec{P}^2}(0)-\lambda_1)+a^4(\pi\rho)(-24 k \sqrt{\vec{P}^2}(\pi\rho)-\lambda_2) \right )
&\nn&
M_{PL}^2 \equiv M^3 R_0 \int dx^5 a^2.
&\eea

In section 3 we explain in detail that if the $P_1$ and/or $P_2$ components of the prepotential are non-zero at the $Z_2$ fixed point then supersymmetry is broken. The reason is that in such a case the BPS solution is not global. Thus unbroken supersymmetry requires $\sqrt{\vec{P}^2}(0)=P_3(0)$ (the same at $x^5=\pi\rho$) and, in consequence, the 4d effective potential vanishes (recall that the brane tensions satisfy (\ref{hyptensions})). 

Similar conclusions hold also in the RS model, with vector multiplets  constructed in ref. \cite{kallosh}. The proof goes the same way as in the hypermultiplet case; the ultimate reason for the vanishing of the effective potential being the supersymmetric tuning of the brane tensions.   

Summarizing this section,  unbroken local supersymmetry in 5d RS-type scenarios implies flat space solutions. This result is somewhat 
unexpected, since within the framework of 4d supergravities we can a priori obtain an anti-de Sitter solution and  preserve supersymmetry at the same time. 
This means that compactifications of the supersymmetric RS scenarios yield a very special subclass of 4d supergravities. This should be kept in mind when  phenomenological models are constructed.  
 
 Another consequence of the vanishing 4d potential is that RS-type models with unbroken supersymmetry cannot  incorporate a mechanism of radion stabilization. The BPS conditions together with the supersymmetric tuning of the brane tensions yield the 4d effective potential which is identically zero. Thus the 4d equation of motion for the radion field is $\frac{\pa M_{PL}^2}{\pa R_0}\bar{R}=0$, which  can be satisfied by any value of $R_0$, since $\bar{R}=0$. 
\section{Supersymmetry breaking and radion 
stabilization in a model with the universal hypermultiplet} 
In the preceding section we argued that RS-type models with unbroken supersymmetry require vanishing of the 4d cosmological constant. At the same time they cannot  incorporate a mechanism of radion stabilization. But  in a realistic model supersymmetry must be broken. Therefore in the remainder of this paper we will study dynamics of the RS model with broken supersymmetry.
 
The RS scenario with spontaneously broken supersymmetry was already discussed in ref. \cite{ls} in the context of radion stabilization. The effective potential for the radion field was generated through the interaction of the radion with gaugino condensates in the bulk.
Communication of supersymmetry breaking  to the visible brane 
occurs at the level of four-dimensional physics with the help of the 
anomaly mediation mechanism proposed in \cite{rs0}.    

In this paper we investigate an alternative mechanism of supersymmetry breaking, similar to that studied in M-theoretical scenarios \cite{horava,ovrut,elpp}. It is triggered by brane sources coupled to the scalar fields in the bulk, which are odd with respect to the $Z_2$ parity. One way to see that supersymmetry is broken is to notice that the Killing spinor cannot be defined globally. The odd fields are the agents that transmit supersymmetry breaking between the hidden and visible branes. Below we present a general description of our mechanism and then apply it to a specific model of 5d gauged supergravity with the universal hypermultiplet. 

Already in the previous section we signalled the possibility of breaking supersymmetry by inducing nontrivial vev of the $P_1$ and/or $P_2$ component of the prepotential. Recall, that the $Z_2$ acts on the supersymmetry parameter $\epsilon^A$ as $\gamma^5 \epsilon^1 (x^5)=\epsilon^1 (-x^5)$, $\gamma^5 \epsilon^2 (x^5)=-\epsilon^2 (-x^5)$. At the $Z_2$ fixed points half of the components are projected out (so that only $\epsilon^1_R$ and $\epsilon^2_L$ are non-zero and they correspond to a single Majorana spinor which generates N=1 supersymmetry in four dimensions).  At the same time, the BPS conditions impose restrictions on the Killing spinors. If we insist on preserving N=1 supersymmetry, the form of the Killing spinors must be consistent with the orbifold projection; in other words, the Killing spinors must have only $Z_2$ even components at the fixed points.

It is straightforward to check that as long as $\cp \sim \sigma_3$ the Killing spinors which generate unbroken N=1 supersymmetry has definite, even  $Z_2$ parity. But as soon as we switch on non-zero $P_1$ or $P_2$  the Killing spinors satisfy the relation:        
\beq
\epsilon^2=\frac{\sqrt{\vec{P}^2}\gamma_5\epsilon^1-P_3\epsilon^1}{P_1-iP_2} .
\eeq
The above equality implies that  if $P_1$ or $P_2$ are non-zero at the $Z_2$ fixed points, then the Killing spinors have not definite parity there and supersymmetry is completely broken.  It is still possible to find a Killing spinor locally, but it cannot be defined globally. Supersymmetry is broken because of the 'misalignment' between the bulk and the brane supersymmetry.

But for $P_1$ or $P_2$ to be non-zero a non-trivial configuration of the $Z_2$ odd fields is necessary. Indeed, from the form of the covariant derivative $
\grad_\alpha\psi_\beta^A + g V_i (P^i)^A_{\;B}\ca_\alpha\psi_\beta^B  
$ it is straightforward to deduce that  while $P_3$ is even with respect to $Z_2$,  $P_1$ and $P_2$ must be odd (recall that $g\sim \epsilon(x^5)$).
 Non-trivial configuration of the odd fields can be induced by sources located on the branes. The mechanism involves coupling the odd scalars  to the branes through their fifth derivatives (which are even and thus well defined on the boundary). The explicit realization of brane sources are  boundary superpotential and boundary gaugino condensates, studied in \cite{flp,mt}. In Appendix B we present a general supersymmetric form of such couplings in 5d supergravity with an arbitrary number of hypermultiplets.  

Studying supersymmetry breaking in 5d supergravity with general hypermultiplet  spectrum is a difficult task. Therefore, from now on we concentrate on a simpler model with only one, so called universal hypermultiplet which includes two even scalars $V,\sigma$ and two odd scalars $\xi, \bar{\xi}$ (the basic properties of the universal hypermultiplet are summarized in Appendix A). 
 One may hope, that the features present in this toy-model persist in more general scenarios, in which supersymmetry breaking is transmitted by a non-trivial configuration of the odd bulk fields. 

 To generate a scalar potential we gauge the isometry $\xi \rightarrow e^{i\phi}\xi$ of the quaternionic manifold \cite{flp2}. Solving the Killing prepotential equation we find:
\beq
P_1=-\frac{{\rm Re}(\xi)}{2\sqrt{V}}\;\;\; P_2=\frac{{\rm Im}(\xi)}{2\sqrt{V}} \;\;\; P_3=\frac{1}{4}(1-\frac{|\xi|^2}{V})  .
\eeq
We see that $P_1$ and $P_2$ are non-zero in this model. Thus, inducing non-zero vev. of the $\xi$ field will break supersymmetry.  To this end, we couple (in a supersymmetric way) the fifth derivative of $\xi$ to sources located on the boundaries.  
The rest of the bosonic action we consider is that given in (\ref{hyperaction}) rewritten for our special choice of the quaternionic manifold and gauging:    
\beqa \label{eq:ac}
&S= M^3\int d^5 x \; e_5 (\frac{R}{2} -\frac{3}{4}(\pa_\alpha\ca_\beta)^2 - \frac{1}{V} |D_\alpha\xi|^2 - \frac{1}{4 V^2} (\pa_\alpha V)^2  + 6k^2(1 + \frac{1}{2 V} 
|\xi|^2 - \frac{1}{2 V^{2} } |\xi|^4 ))& \nonumber \\
&-M^3\int d^5 x \; \frac{e_5}{e^{5}_{5}} (\delta(x^5)-\delta(x^5-\pi\rho)) 6k (1-\frac{|\xi|^2}{V}) &\nn&
 -\int d^5 x \; e_5 \frac{2}{V g_{55}}\left( 
\delta(x^5)W_1(\pa_5 \xi + 2 \delta(x^5) \frac{\bar{W}_1}{M^3})    
+ \delta(x^5-\pi\rho)
W_2 ( \pa_5 \xi +2 \delta(x^5-\pi\rho) \frac{\bar{W}_2}{M^3})+ h.c \right).
&\nn&&
\eeqa   
 In the first line we displayed the relevant kinetic terms (we neglected the field $\sigma$ which does not play any role in the following). The covariant derivative acting on $\xi$ is $D_\alpha \xi = \pa_\alpha \xi + i g\xi \ca_\alpha$ with the gauge charge $g=3k\sqrt{2}\epsilon(x^5)$. The boundary terms in the second line are proportional to $P_3$ and are necessary for local supersymmetry of the 5d action \cite{flp2}. The third line contains derivative coupling of the odd field to the boundary sources $W$; the presence of singular $\delta^2$ terms is  commented on in Appendix B. For the sake of concreteness 
we have concentrated on the case, where sources of supersymmetry breakdown are represented by expectation values of brane superpotentials (gaugino condensates will  be discussed later in the paper).

Locally in the bulk, it is possible to find a flat BPS solution which preserves half of the supersymmetry (see Appendix A for details), but as soon as we switch on non-zero sources $W$ the BPS solution does not satisfy the matching conditions at the $Z_2$ fixed points. In such case we search  for maximally symmetric, non-BPS solutions.        
 We take the ansatz that allows for $adS_4$ foliation of the 4d metric, 
$g_{\alpha \beta} = \; diag \, (-a^2 (x^5 ) e^{2 L x^3}, a^2 (x^5 ) e^{2 L x^3}, a^2 (x^5 ) e^{2 L x^3}, a^2 (x^5), R_0^2)$ (the generalization to the de Sitter  foliation is straightforward); the size of the 5th dimension is parametrized by $R_0$. Our vacuum has the form of a constant curvature foliation (otherwise it could lead to violation of Lorentz invariance after integrating 
out the extra dimensions \cite{grojean}).
 The ansatz for the scalar field and the graviphoton is:
\beq
\xi=\xi(x^5) \;\; V=V(x^5) \;\; \sigma=const \;\; \ca_\mu=0 \;\; \ca_5=const
\eeq
We start with the RS solution $a=e^{-k R_0|y|},\;\xi=0$ and treat the boundary sources as a perturbation. This procedure is justified since we expect that the parameters of the bulk lagrangian, which determine the zeroth-order solution,  are close to the Planck scale, while the boundary sources should be of the order of supersymmetry breaking scale.  Since the sources set the boundary value of $\xi$ we expect that $\xi$ will get $(\frac{W}{M^3})$ corrections. 
On the other hand, the $\xi$ enters quadratically into the equation for the metric, so the metric will get the correction only at the order $(\frac{W}{M^3})^2$. We will be able to find a solution  valid to the order $(\frac{W}{M^3})^2$ . First we consider the linearized equation of motion for the $\xi$ field which  has the form:
\beqa \label{eq:xi} 
& \xi''+  \xi' (4 \frac{a'}{a}+2ig \ca_5)+ 
\xi (3 k^2 R_0 -4i g k R_0 \ca_5 - g^2 \ca_5^2) = 
-2 \delta' (x^5) \frac{\ov{W}_1}{M^3} - 2 \delta' (x^5 - \pi \rho ) 
\frac{\ov{W}_2}{M^3} 
&\eeqa
The solution is: 
\beq
\xi = C \epsilon(x^5) e^{k( R_0 -  3\sqrt{2}i\ca_5)|y|} 
\eeq
Matching the $\delta'$ in the equation of motion yields the boundary conditions: \\
$\xi(0^+)=-\frac{\ov{W}_1}{M^3}, \;\;\xi(\pi\rho^-)=\frac{\ov{W}_2}{M^3}$. 
As a consequence:
\bea \label{bc:w}
&C= -\frac{\bar{W}_1}{M^3}, \;\;\; C e^{( R_0 -  3\sqrt{2}i\ca_5)k\pi\rho}=
\frac{\bar{W}_2}{M^3}.&
\eea
In the absence of supersymmetry breaking the moduli $R_0$ and $\ca_5$ could have arbitrary constant values. When we switch on the sources for the odd fields  and break supersymmetry, the expectation value of the moduli is determined  by the boundary sources $W_i$. In other words, the moduli are stabilized. Hence, in our model the supersymmetry breaking indeed 
leads to stabilization of the fifth-dimension.    In the 4d effective theory $R_0$ and $\ca_5$ enter the so called $T$ supermultiplet. 
In the next section we show, that the  configuration of the  $\xi$ field which we have found gives rise to the 4d effective potential which is able to stabilize $T$ modulus when the 4d dilaton is frozen.

The non-zero value of $\xi$ of order $W$ will produce a back-reaction on the metric and the scalar $V$ of order $W^2$. Keeping only terms relevant to the  order $W^2$, the Einstein equations read:     
\beqa \label{einstein} 
&3 \frac{a''}{a}+ 3 \left ( \frac{a'}{a} \right )^2 +3 \frac{L^2}{a^2}
+ \frac{1}{V}|\hat{D_5}\xi|^2  - 6 k^2 ( 1 
+ \frac{1}{2 V} |\xi|^2 )
= -(\delta (x^5)- \delta (x^5 - \pi \rho)) 6 k (1-\frac{|\xi|^2}{V})    
&
\nn
&6 \left ( \frac{a'}{a} \right )^2 +6 \frac{L^2}{a^2}    
- \frac{1}{V}|\hat{D_5}\xi|^2- 6 k^2 ( 1 
+ \frac{1}{2 V} |\xi|^2 )= 0 
&\eeqa
The terms involving boundary sources $W$ gather nicely into the full square with the kinetic term of $\xi$. The boundary conditions (\ref{bc:w}) ensure that the hatted derivative contains no delta functions . Thus the only boundary terms  we have to consider are those displayed explicitly in (\ref{einstein}). Matching the delta functions in the first equation yields the boundary conditions for the warp factor:
\bea
\frac{a'}{a}(0)=-6k(1-|\xi(0)|^2) & \frac{a'}{a}(\pi\rho)=-6k(1-|\xi(\pi\rho)|^2)&
\eea   
Plugging in our solution for $\xi$ we can satisfy the boundary conditions if the warp factor has the form:
\beq \label{eq:warpnew}
a(x^5)=e^{-kR_0|x^5|} + \frac{|C|^2}{2V}e^{kR_0|x^5|}
\eeq
 Away from the branes the Einstein equations can be satisfied if we choose $L^2=\frac{16}{6}\frac{k^2|C|^2}{V}$. This means that the 4d curvature is $\bar{R}=-32\frac{k^2|C|^2}{V}$, hence this solution corresponds to anti-de Sitter 
4d foliation.

Similarly, the scalar $V$ will get the correction of the order of $(\frac{W}{M^3})^2$. The equation of motion is:
\beq
V'' + \frac{4a'}{a}V' + 2|\hat{D}_5 \xi|- 6 (k R_0)^2 |\xi|^2 = 
12 k R_0 |\xi|^2(\delta(x^5)-\delta(x^5-\pi\rho))  
\eeq 
 which is solved by  $V=V_0 -|C|^2 e^{2kR_0 |x^5|}$. 

Summarizing, we have found a perturbative (to the order $|W^2|$) solution to the equation of motion in the presence of non-zero boundary sources  for the odd field $\xi$:
\bea&
\label{solutionall}
\xi = C \epsilon(x^5) e^{k( R_0 -  3\sqrt{2}ik\ca_5)|x^5|}, 
&C= -\frac{\ov{W}_1}{M^3},\;\;\;\;Ce^{k( R_0 -  3\sqrt{2}ik\ca_5)\pi\rho}=\frac{\ov{W}_2}{M^3},\nn&
 a=e^{-kR_0|x^5|} + \frac{|C|^2}{2V_0}e^{kR_0|x^5|},&
 L^2=\frac{8}{3}\frac{k^2|C|^2}{V_0},  
\nn
&V=V_0 -|C|^2 e^{2kR_0 |x^5|},& \sigma=\sigma_0.  
\eea
In the above solution $V_0$ and $\sigma_0$ are  arbitrary constants. Hence these moduli are not stabilized.  In fact, matching conditions in the equation of motion for $V$ require $V_0 \rightarrow \infty$. This means that $V_0$ exhibits the runaway behaviour.  To achieve stabilization of all the moduli we 
need to complicate our model by adding boundary sectors that couple to $V$.   
  
As a cross-check of the above results we can calculate the 4d effective potential, obtained by integrating out the 5d bosonic action in the background (\ref{solutionall}). The result (to the order $\frac{W}{M^3}^2$) is:
\beq
\cl_4 = \sqrt{-\bar{g}}\frac{M^3}{k}(1-e^{-2kR_0\pi\rho})(\frac{1}{2}\bar{R} + 8\frac{k^2|C|^2}{V_0}).
\eeq
We denoted by $\bar{g}$ the oscillations of the 4d metric around the vacuum solution. Solving the Einstein equations in the 4d effective theory yields $\bar{R}=-32\frac{k^2|C|^2}{V_0}$ which is consistent with the value of $L^2 \equiv -\frac{1}{12}\bar{R}$ in (\ref{solutionall}). We also see that $V_0$ enters the denominator of the effective potential, which explains its runaway behaviour commented on earlier.
      
Before closing the discussion of the 5d classical solutions 
and supersymmetry breakdown, let us comment on proposals \cite{verlinde, schmid,pmayr} of solving the 
cosmological constant problem due to supersymmetry of the bulk-brane 
system. 
To put the issue into the perspective, let us note that the second equation 
in (\ref{einstein}) does not contain second 
derivatives of fields, hence it acts as a sort of constraint on the 
solutions of the remaining equations. This becomes more clear in the Hamiltonian 
approach towards the flow along the fifth dimension, where this equation 
arises as the hamiltonian constraint ${\cal H} =0$, and is usually 
used to illustrate the way the conservation of the 4d curvature $ L^2$ 
is achieved through the compensation between gradient and potential terms 
along the classical flow. 
However, this classical conservation hinges upon 
fulfilling certain consistency conditions between brane sources, or 
between boundary conditions induced by them, as illustrated by the model 
above. When one perturbs the boundary terms on one wall, then 
to stay within the family of maximally symmetric foliations one of two things 
must happen. Either the distance between branes must change, or the 
source at the distant brane must be retuned. In the class of models 
which we constructed, if the 4d curvature is present then supersymmetry is broken , and doesn't take care of such a retuning. Furthermore, even if 
retuning takes place,
the size of 4d curvature, i.e. of the effective cosmological constant,
does change as well; moreover, the magnitude of the effective cosmological 
constant has quadratic dependence on the boundary terms 
which induce supersymmetry breakdown. Hence, any perturbation of the 
boundary, instead of being screened by the bulk physics, contributes  
quadratically to the effective cosmological constant. 
Of course, we are talking about perturbations that can be considered 
quasi-classical on the brane.    
Thus we do not see here any special new effect of the extra dimension 
in the cancellation of the cosmological constant. The positive 
aspect of supersymmetry is exactly the one we know from 4d physics.
Supersymmetry, even the broken one, limits the size of the brane terms 
inducing supersymmetry breakdown, thus limiting the magnitude 
of the 4d cosmological constant, since the two effects are strictly related 
to each other. \\

\section{Four-dimensional effective theory}

In this section we give the form of the effective four-dimensional 
 supergravity describing zero-mode fluctuations in the model presented in the previous section. Since in the 5d set-up supersymmetry was broken spontanously, it is safe to  assume 
that in 4d this supersymmetry breakdown can be considered as a spontaneous 
breakdown in a 4d supergravity Lagrangian described with the help of 
certain K\"ahler potential $K$, superpotential $W$ and gauge kinetic functions
$H$. 
The goal is to identify reliably these functions starting from the 
maximally symmetric approximate solutions (\ref{solutionall}) we have found in the previous section.  
Our procedure is perturbative in the supersymmetry breaking parameter $W_1$. 
The configurations (\ref{solutionall}) are solving equations of motion and boundary 
conditions to the second order in $\frac{W_1}{M^3}$. However, it is  sufficient 
to identify the functions we are looking for from the terms 
which can be reliably read at the order $(\frac{W_1}{M^3})^1$. Such terms include 
the gravitino mass term. In addition, we have 
at our disposal the complete kinetic terms for moduli, gauge and matter fields,
which are of order $(W_1)^0$ and are sufficient to read off the K\"ahler 
potential for moduli and matter fields. The complete procedure consists 
 of solving to the given order for all background fields, including the 
$Z_2$-odd ones, substituting the solutions back to the 5d Lagrangian and 
integrating over the fifth dimension. This procedure can be carried out
to the full extent, 
however here, taking the existence of the effective 4d supergravity for granted, 
we shall perform the integration only for certain relevant terms - the ones 
which give direct information about $K$, $W$ and $H$. 

\subsection{K\"ahler function for the radion and 4d dilaton}

To the lowest order in field fluctuations the ansatz defining the radion 
modulus in the Randall-Sundrum background, which is also the lowest order 
solution in $W_1$-expansion, is 
\beq \label{eq:ans}
ds^2 =  e^{-k R_0 (x^\mu) |x^5| } \bar{g}_{\mu\nu}dx^\mu dx^{\nu} + R_{0}^2 (x^\mu) 
(dx^5)^2.
\eeq
{ This ansatz leads to the K\"ahler potential:
\bea \label{eq:ourk}
&K=-M_P^2\log(S + \bar{S}) -3 M_P^2 \log \left ( f(T+ \bar{T}) \right ) &
\nn
&S=V_0 + i \sigma_0, \;\;\;\;\; T= k\pi\rho (R_0 + i\sqrt{2}\ca_5),& 
\eea
where we defined $M_P^2= \frac{M^3}{k}(1-e^{-2k\pi\rho\langle R_0 \rangle})$, $f= \beta (1- e^{-(T + \bar{T})})$ with 
$\beta = \frac{ M^3}{k M^{2}_{P}}$. The form of the K\"ahler potential for the multiplet T was previously derived in \cite{bagger2, ls}.  

To see in more detail how the argument goes, let us summarize those 
terms in the 
5d action that are most relevant to the forthcoming discussion:
\beqa \label{eq:relgravdil}
&S_{grav}= M^3 \int d^5xe_{5} \left (\frac{1}{2}R
-\frac{1}{2}\ov{\psi_\alpha}^A\gamma^{\alpha\beta\gamma}D_\beta\psi_{A\gamma} 
-3 i k\epsilon(x^5)  \ov{\psi_{\alpha}}^A\gamma^{\alpha\beta}
\psi^{B}_{\beta}{\cal P}_{AB} \right )  & \nonumber \\
&S_{hyp}=M^3\int d^5xe_{5} \left ( \right .   
-\frac{1}{4V^2}(\pa_{\alpha}V\pa^{\alpha}V+\pa_{\alpha}\sigma \pa^{\alpha}\sigma) -\frac {1}{V}\hat{D}_{\alpha}\xi \hat{D}^{\alpha}\bar{\xi}
  -\frac{1}{2}\ov{\lambda}^a\gamma^\alpha D_\alpha \lambda_a &\nn&
  -\frac{1}{2} \ov{\psi_\alpha}^A\gamma^{\alpha\beta\gamma}\psi^B_{\gamma}
 \omega_{uAB}\hat{D}_5 q^u   
\left . \right),
&\eeqa 
where $\hat{D}_5 \xi = \pa_5\xi +3i\r2 k\epsilon(x^5)\xi\ca_5 + 2\delta(x_i)\bar{W}_i$   , the prepotential is given by $\cp=
(\frac{1}{4}- \frac{x^2+y^2}{4V}) i \sigma^3 - \frac{x}{2V^{1/2}} i\sigma^1 
+ \frac{y}{2V^{1/2}} i\sigma^2 ) $
 and the $SU(2)$ spin connection $\omega$ is given in  Appendix A.

The goal is to reduce various terms in this action down to the Einstein 
frame in four dimensions, where the matter-supergravity action is of  the 
standard form of Cremmer et al. \cite{cremmer,wess-bagger}. 
After inserting the ansatz (\ref{eq:ans}) into the 5d gravitational action,  substituting the 5d gravitino with the 4d zero-modes $\psi_{\mu R}^1= a^{1/2} (\psi_{\mu R})_{4d}\;\;\;\psi_{\mu L}^2= a^{1/2} (\psi_{\mu L})_{4d} $ (see \cite{flp2,gp2}),   
and integrating over $x^5$ one finds the following graviton and gravitino kinetic terms 
in 4d: 
\beq
\frac{1}{2} M^{2}_{P} {e}_4 \beta (1-e^{-2 k\pi \rho R_0}) \tilde{R}^{(4)}
-\frac{1}{2} M^{2}_{P} {e}_4 \beta (1-e^{-2k\pi \rho R_0})
\ov{\psi}_\mu \gamma^{\mu \nu \rho} D_\nu \psi_\rho + ...
\eeq
where the $R_0$ dependent bracket is simply the scalar part of the 
real vector superfield $f$ (in what follows we shall use the symbol $f$ for this scalar function as well).
After performing the Weyl rescaling of the graviton and gravitino 
\beq \label{eq:weyl}
e^{a}_\mu \rightarrow e^{a}_\mu f^{-1/2}, \;\; \psi_\mu \rightarrow f^{-1/4}, 
\psi_\mu
\eeq
one arrives at the canonical action for graviton and gravitino in 
four dimensions. Substituting the ansatz (\ref{eq:ans}) into the 5d 
action results also in (a part of) the kinetic terms for $R_0$
$\delta S_{kin} = - 3 M^{3} k (\pi\rho)^2 e^{-2kR_0\pi\rho} \int d^4 x e_4  
\pa_\mu R_0 \pa^\mu R_0 $};
together with another term with two derivatives on $ R_0 $ generated by the Weyl 
rescaling, this gives exactly the kinetic energy reproduced by the K\"ahler 
potential $K$ given in (\ref{eq:ourk}). 
A procedure of rescalings and integrating over $x^5$ applied to the 
kinetic term of $V$ (and $\sigma$) also produces 4d kinetic terms which are  
immediately seen to be exactly given by (\ref{eq:ourk}).
It is worth noticing at this point that, contrary to assumptions usually made,
the natural 5d frame where we have coupled bulk gravity and moduli 
with general gauge sectors on branes {\it does not} directly give us 
the 
superspace frame of the 4d supergravity. This is seen from the fact, that
the Weyl rescaling in 4d {\it is not} given by the complete K\"ahler 
function, but only by the part of it that depends on the radion.
Fortunately, radion and hypermultiplet moduli do not have a kinetic mixing 
in the canonical 5d frame.

\subsection{Effective superpotential}

Knowing the K\"ahler function and performing the reduction of 
the gravitino mass terms (the ones which in five-dimensions couple $Z_2$ even components with 
even components) one can identify the effective 4d superpotential
\beq \label{eq:oursup}
W= 2 \sqrt{2} (W_1 + e^{-3 T} W_2 ).
\eeq

One can see that the contribution to the effective superpotential given by 
the second brane at $x^5=\pi\rho$ is suppressed by the factor $a^{3}_0$ with respect to the 
contribution from the Planck brane. This fits nicely the notion of a  
universal down-scaling of all the mass scales on the visible brane.   
 
The derivation  of the superpotential goes as follows. There are three contributions to  the 4d gravitino mass terms:
\beq
\cl_{gm}=M^3 e_5 \frac{1}{\sqrt{V}} \left ( -\frac{1}{2}\hat{D}_5 \xi  \ov{\psi^{1}_\mu} \gamma^{\mu \nu}\gamma^5 \psi^{2}_\nu  
-\frac{3}{2}k\epsilon(x^5)\xi\ov{\psi^{1}_\mu} \gamma^{\mu \nu} \psi^{2}_\nu 
+i\ca_5\frac{3}{2\r2}k\epsilon(x^5)\xi\ov{\psi^{1}_\mu} \gamma^{\mu \nu} \psi^{2}_\nu 
 +h.c \right)
\eeq
The first  contribution comes from the term in (\ref{eq:relgravdil}) involving the $\xi$ part of the spin connection $\omega$, the second from the 5d gravitino mass term involving the prepotential $\cp$. Finally the term with the graviphoton $\ca_5$ comes from the SU(2) covariant derivative in the gravitino kinetic term:
$D_\alpha \psi_\beta^{A} = \nabla_\alpha \psi_\beta^A +g \ca_\beta P^A_{\;\;B} \psi_{\beta}^{B}$.
Plugging in the gravitino zero modes in the solution (\ref{solutionall}), 
integrating over $x^5$ and Weyl rescaling, we get: 
\bea
&\cl_{4dgm}= \frac{1}{2\sqrt{V_0}f^{3/2} M^{2}_P} e_4 \bar{W}_1 (1-e^{-k\pi\rho(2R_0 +3\r2i\ca_5)})
\ov{\psi}_{\mu R}\gamma^{\mu\nu}\psi_{\nu L} +h.c
=&\nn&
\frac{1}{2\sqrt{V_0} f^{3/2} M^{2}_P} e_4 (\bar{W}_1 +e^{-3k\pi\rho R_0}\bar{W}_2)
\ov{\psi}_{\mu R}\gamma^{\mu\nu}\psi_{\nu L} +h.c
&\eea
In the last step we used the boundary condition (\ref{bc:w}). In the standard 4d supergravity formulation \cite{cremmer} this term has the form $\frac{e_4}{2}e^{G/2}\ov{\psi}_{\mu R}\gamma^{\mu\nu}\psi_{\nu L} +h.c$ with $G=K +\ln|W|^2$. The form of the K\"ahler potential (\ref{eq:ourk}) and the holomorphicity of the superpotential requires the superpotential of our 4d model to be exactly (\ref{eq:oursup}).    
 
The non-trivial test of the consistency of the superpotential (\ref{eq:oursup}) comes from minimizing the 4d effective scalar potential derived from (\ref{eq:ourk}) and (\ref{eq:oursup}). In the standard formulation it has the form: ${\cal V} = e_4 e^{G}(G_iG^{i\bar{j}}G_{\bar{j}}-3 )$
 or explicitly:
\bea&
{\cal V}= e_4 \frac{4}{M^{2}_P V_0 \beta^3 (1-e^{-2k\pi\rho R_0})^3} \left(\right.
|W_1|^2 (3 e^{-2k\pi\rho R_0}-2)
+ |W_2|^2 (3 e^{-4k\pi\rho R_0} - 2 e^{-6k\pi\rho R_0})
&\nn&
+W_1\bar{W_2}e^{-3k\pi\rho{R_0-i\r2\ca_5}}
+W_2\bar{W_1}e^{-3k\pi\rho{R_0+i\r2\ca_5}}
\left. \right)
&\eea
 
Minimizing the above scalar potential with respect to $R_0$ and $\ca_5$ yields the relation (\ref{bc:w}), consistently with the 5d picture. Also, the value of the prepotential  at the minimum (that is the cosmological constant) is consistent with the solution (\ref{solutionall}). Note also that, the 4d effective potential is of the runaway type with respect to $V_0$.
 
Just as a remainder one should mention that the usual redefinition of the 4d 
components of the gravitino leading to removal of the kinetic mixing terms 
between  
$\psi^{A}_\mu$ and $\psi^{B}_5$ is needed $\psi^{A}_\mu \rightarrow 
\psi^{A}_\mu + \frac{1}{2 e^{5}_{5}} \gamma_\mu \gamma^5 \psi^{A}_5$. 
Inspecting more closely the fermionic mass matrix one notices readily 
the mass terms proportional to $\langle \pa_5 \xi \rangle $ and 
$\langle \xi \rangle$ which mix 4d gravitini with $\psi_5$ and hyperini $\lambda$. 
The origin of these mass terms is analogous to that of gravitini masses,
and their presence signals that the superhiggs mechanism is at work 
(as expected). To see that the higgsino is a mixture of hyperino and 
modulino $\psi_5$ one can inspect the supersymmetry transformation laws
of these fermions:
\beqa
&\delta \psi^{1}_{5}  = -\frac{1}{\sqrt{V}} \pa_5 \xi \epsilon^2 + \delta_{i}(x^5)
\frac{2 i e^{5}_{5}}{\sqrt{V}} W \epsilon^2+ i 2 k \epsilon(x^5)
 (-\frac{{\rmfamily Re }(\xi) }{2 \sqrt{V}} (\sigma^{1})^{1 B} +\frac{{\rmfamily Im } (\xi)}{2 \sqrt{V}} (\sigma^{2})^{1 B}) 
 \gamma_5 \epsilon_B & \nonumber \\
&\delta \psi^{2}_{5} = +\frac{1}{\sqrt{V}} \pa_5 \bar{\xi} \epsilon^1 - \delta_{i}(x^5)
\frac{2 i e^{5}_{5}}{\sqrt{V} } \bar{W} \epsilon^1+ 
i 2 k \epsilon(x^5) 
(-\frac{{\rmfamily Re }(\xi)}{2 \sqrt{V}} (\sigma^{1})^{1 B} +\frac{{\rmfamily Im } (\xi)}{2 \sqrt{V}} (\sigma^{2})^{1 B})
\gamma_5 \epsilon_B
& \nonumber \\
&\delta \lambda^1= +\frac{i}{\sqrt{2V}} \gamma^5 \pa_5 \xi \epsilon^2 - 
\delta_{i}(x^5)
\frac{2 \sqrt{V}}{a} \bar{W} \epsilon^2+ 3 k \epsilon(x^5)
V^{A1}_u k^u  \epsilon_A
& \nonumber \\
&\delta \lambda^2= +\frac{i}{\sqrt{2V}} \gamma^5 \pa_5 \bar{\xi} \epsilon^1 - 
\delta_{i}(x^5)
\frac{2 \sqrt{V}}{a} \bar{W} \epsilon^1 +
3 k \epsilon(x^5)
V^{A2}_u k^u \epsilon_A
&
\eeqa
just repeating the procedure given in \cite{flp} for the Horava-Witten model.
In the above
$V_{u}^{Ab}$ are  $SU(2)$ vielbeins given in Appendix A. Since the theory has a mass gap $\Delta m = m_{KK} \approx M a(\pi \rho )$, 
to find the low-energy goldstino one needs to substitute into above equations 
the vacuum solutions for bulk scalars and to project resulting expressions onto their zero-mode components.  

Another way to identify the 4d goldstino 
is to derive $\langle F_S \rangle$ and $\langle F_T \rangle$ with the help of the 4d effective Lagrangian
and we shall give the result at the end of this chapter.

\subsection{Gauge kinetic functions} 
To arrive at a realistic model, one should introduce gauge and charged matter fields. One option --- with gauge fields living in five-dimensional vector multiplets --- was studied in \cite{ls,gp2} and yields the gauge kinetic function $H \sim T$. Here we present an alternative, 
a model with gauge and matter fields confined on the boundaries. 
The action of the gauge sector is:      
\beqa \label{eq:syms}
&S_{YM}= \int d^5x \frac{e_5}{e_5^5}\delta(x^5) 
\left (
 -\frac {V} {4} 
F_{\mu\nu}^{a} F^{a\mu\nu}   
-\frac{1}{4} 
\sigma F_{\mu\nu}^{a} \tilde F^{a\mu\nu} 
-\frac {V} {2} 
\overline{\chi^{a}}D\!\!\!\!\slash\chi^{a} 
+\frac{V}{4}  
(\overline{\psi}_{\mu} \gamma^{\nu\rho}\gamma^{\mu}\chi^{a})F^{a}_{\nu\rho} 
\right . & \nonumber \\
&+\frac {3i} {4\sqrt{2}}\frac {V}{e_{5}^{5}} 
(\overline{\chi}^{a}\gamma^{5}\gamma^{\mu}\chi^{a}){\cal F}_{\mu 5}
-\frac {1} {4} 
(\overline{\lambda}\gamma^{\nu\rho}\chi^{a})F_{\nu\rho}^{a} 
-\frac {i} {8} 
(\overline{\chi}^{a}\gamma^{5}\gamma^{\mu}\chi^{a})\partial_{\mu}\sigma 
& \nonumber \\
&\left . -\frac{\sqrt{V}}{2e_{5}^{5}} 
( 
(\overline{\chi^{a}}_{L}\chi^{a}_{R})\partial_{5}\overline{\xi} 
+ (\overline{\chi^{a}}_{R}\chi^{a}_{L})\partial_{5}\xi 
) 
+({\rmfamily 4-fermi}) 
  \right )& 
\eeqa
and similarly on the visible brane at $\pi\rho$. 
In the above the bulk fermions appear in their even (and Majorana in the 4d sense) combinations defined as:  
\bea 
\label{spinors}  
 \psi_{\mu} = \left ( 
\begin{array}{cccc} 
{\psi_{L\mu}^{2}} \\ 
 {\psi_{R\mu}^{1}} 
\end{array} 
\right ), 
& 
 \lambda =\sqrt{2}V \left ( 
\begin{array}{cccc} 
{-i\lambda_{L}^{1}} \\ 
 {i\lambda_{R}^{2}} 
\end{array} 
\right ),  
\eea   
and brane supersymmetry is generated by the Majorana fermion 
$\epsilon = \left ( 
\begin{array}{cccc} 
{\epsilon_{L}^{2}} \\ 
 {\epsilon_{R}^{1}} 
\end{array} 
\right ) $.
 The  tree-level 
gauge kinetic function turns out to be universal and equal:
\beq
H(S)=S
\eeq
on either brane. However, we expect corrections to this universality 
(see the forthcoming discussion). 

\subsection{K\"ahler function for matter fields}

Now we introduce the matter fields living on the branes. We allow the superpotentials $W$ to depend on the scalar $\Phi$. Let us concentrate on the matter living on the visible brane at $x^5 = \pi \rho$.  The  part of the action relevant to our discussion is (more terms and corrections to fermionic transformations 
are given in Appendix C):
\beq \label{eq:mbrane}
L_{m, \, brane} = \frac{e_5}{\sqrt{g_{55}}} \delta(x^5-\pi\rho) \left (-D_\mu \Phi D^\mu
\bar{\Phi} - \frac{4}{V} \frac{\partial W_i}{\partial \Phi} 
\frac{\partial \bar{W}_i}{\partial \bar{\Phi}}+ ...\right ).
\eeq
In passing from the canonical 5d frame to the 4d Einstein frame 
one needs two rescalings of the metric. First, one factorizes out of the original metric the warp factor and, second, one performs the Weyl rescaling 
$e^{a}_\mu \rightarrow e^{a}_\mu f^{-1/2}$. 
The second, potential term in the brane Lagrangian (\ref{eq:mbrane}) 
does not get corrected through the Weyl rescaling of the curvature scalar $R$ 
(all new terms borne this way carry two space-time derivatives);  
 after rescalings it becomes 
\beq \label{eq:4dpot}
L_{pot, \, b} = 4e_4 \frac{1}{V_0} f^{-2} e^{-4 k \pi \rho R_0}
  \left |\frac{\partial W_2}{\partial \Phi} \right |^2.
\eeq
The canonical 4d SUGRA expression for such a term is
$ -e_4 e^K g^{\Phi \bar{\Phi}} |\partial_\Phi W|^2$, where $g^{\Phi \bar{\Phi}}$ is the 
inverse K\"ahler metric for the matter fields. Comparing these two 
expressions, one obtains information about the matter K\"ahler metric. In addition, one should remember that in the limit $k \rightarrow 
0$  the matter kinetic terms are reproduced 
with the K\"ahler function $K_0 = -\log(S + \bar{S}) -3 \log(T + \bar{T} -
\gamma |\Phi|^2)$ with a suitable coefficient $\gamma $.
This suggests a trial function of the form 
\beq \label{eq:kah}
K(T,\bar{T};\Phi,\bar{\Phi}) = -M_P^2\log(S + \bar{S}) -3 M_P^2\log \left ( f(T+ \bar{T}
-\gamma |\Phi|^2 ) \right ) 
\eeq
where $f= \beta (1- e^{-(T + \bar{T} -\gamma |\Phi|^2 )})$.   
And indeed, the term (\ref{eq:4dpot}) is reproduced upon substituting 
$K^{\Phi \bar{\Phi}} |_{\Phi=0}$ into the standard supergravity expression with 
$\gamma=\frac{k}{3M^3}$. In addition we need to redefine the real part of 
the $T$ modulus: ${\rmfamily Re} T = R_0 - \frac{\gamma}{2}|\Phi|^2$.
   
The additional check comes from the first term in (\ref{eq:mbrane})
which after rescalings takes the form $-e_4 f^{-1} e^{-2k 
\pi \rho R_0} D_\mu \Phi D^\mu
\bar{\Phi}$. In canonical supergravity it should equal $ -e_4 \frac{\partial^2 K}
{\partial \Phi \partial \bar{\Phi}}  D_\mu \Phi D^\mu \bar{\Phi}$ 
and indeed it is for our choice (\ref{eq:kah}) with $\gamma=k/3M^3$
(one should notice that $f_T |_{\Phi=0} = e^{-2k\pi \rho R_0}$).

The above discussion applies to matter living on the warped brane (whenever 
necessary we should denote it by $\Phi_2$). To include matter living on the 
Planck brane ($\Phi_1$) we need to improve the K\"ahler potential further. 
The expression which reproduces properly also matter Lagrangian on 
the first, unwarped, brane is 
\beq
K(\Phi_1,\Phi_2) = -M^{2}_P \log(S + \bar{S}) -3 M^{2}_P \log 
\left ( f(T+\bar{T} - \gamma |\Phi_2|^2)
-\beta \gamma |\Phi_1|^2 \right )
\eeq
where as before $\gamma = \frac{k}{3 M^3}$. 

To summarize, we have deduced the zeroth-order approximmations to the 
K\"ahler function, superpotential and gauge kinetic functions of the 4d 
effective supergravity for 5d warped Randall-Sundrum model with general 
gauge and matter sectors on the branes. \\
It turns out that the leading effects of the Randall-Sundrum brane tensions 
are encoded in the exponential dependence of the effective K\"ahler function 
on radion and matter fields, and in the exponential suppression of the 
contributions to the effective superpotential borne on the warped brane. 

\subsection{Gaugino condensates}

In this context it is interesting to ask the question about the proper 
immersion of the gaugino condensates into the effective supergravity 
picture. Let us discuss this issue at the level of the 4d 
model we have just constructed. 
The basic expression for the effective potential including the 
contribution from the gaugino condensates we start with is 
\beq \label{eq:potlam}
V=e^K g^{S \bar{S}} | D_S W + \frac{1}{4} e^{-K/2} \langle \bar{\lambda} \lambda \rangle |^2 .
\eeq
Usually, for canonically normalized gauge and gaugino fields, one replaces 
the condensate by $\Lambda_{c}^3= M^{3}_{GUT} e^{-\frac{3 {\rmfamily Re}(S)}{2 b_0}}$. 
Using holomorphicity and R-symmetry of the gauge sector one often promotes 
this contribution to the one generated by an effective superpotential 
for the dilaton superfield $S$.     
However, the question arises of 
what we shall substitute for $M_{GUT}$ in the warped case.
To answer this we should carefully recompute the condensation scale 
$\Lambda_c$ using the one-loop renormalization group equation for the gauge 
coupling and watching the rescalings we make before reaching the 4d 
canonical frame. The point is that, to achieve canonical normalization of 
gauginos, we must perform the rescaling $\lambda \rightarrow a^{-3/2} f^{3/4} 
\lambda$. These rescalings are anomalous and amount to threshold 
corrections at the 4d upper scale of running which is $M_P$
\beq \label{thresholds} 
\left ({\rmfamily Re}(S) = \right )\frac{1}{g^{2}(M_P)} \rightarrow {\rmfamily Re}(S) + 
(\frac{1}{2} \log f - \log a ) 2 b_0.
\eeq
Now the renormalization group running looks like 
\beq
 \frac{1}{g^{2}(p)}={\rmfamily Re}(S) + (\frac{1}{2} \log f - \log a )
2 b_0 - 2 b_0 \log(\frac{M_P}{p}).
\eeq
This gives the condensation scale 
\beq
\Lambda^{3}_c = M_{P}^3 a(\pi \rho)^3 f^{-3/2} e^{-\frac{3 {\rmfamily Re}(S)}{2 b_0} }. 
\eeq
Thus we see that the effective $M_{GUT}$ at the warped brane equals 
essentially $M_P a(\pi \rho)$ (but $M_{GUT}=M_P$ on the Planck brane). 
We can substitute the condensation scale that we have just derived into the potential (\ref{eq:potlam}) to obtain
\beq \label{eq:finpot}
V= e^K (S +\bar{S})^2 | D_S W + \frac{1}{4} (S + \bar{S}) M^{3}_P e^{-3k\pi \rho R_0 } e^{-\frac{3 {\rmfamily Re}(S)}{2 b_0} }|^2 
\eeq
(where by $W$ we mean a perturbative superpotential plus possible constant contributions inherited from higher dimensions).

It is tempting to replace the contribution from the condensates by the effective superpotential 
\beq \label{wnpert}
W_{npert} =M^{3}_P e^{-3 T} e^{-\frac{3 S}{2 b_0} } 
\eeq
(the factor ${\rmfamily Re}(S)$ multiplying the 
exponential dependence on $S$ can safely be neglected). 
This would fit nicely with 
the formula for the perturbative 4d superpotential. However, one should 
notice that plugging such a superpotential into the standard SUGRA 
expression would generate in the potential new terms which are not present   
in the expression (\ref{eq:finpot}), namely these containing $D_T W_{npert}$,
which would be nonvanishing. 
This inconsistency is alleviated if one notices that together with threshold 
corrections we have used, there are suitable 
`one-loop' corrections to gauge kinetic function on the warped brane. 
The threshold correction on that brane may be split into two pieces.
The first is the same as on the Planck brane and comes from the Weyl rescaling 
by the power of $f$. The $f$ became identified with a scalar component of a 
real superfield defining 
the K\"ahler potential, hence this part of the $1$-loop corrections
 is fully analogous 
to result of the Weyl rescaling leading from superspace to canonical frame 
given by Bagger et al. \cite{Bagger:2001dh}. 
The second piece is nonuniversal, and is associated with additional powers of 
the warp factor $a$ multiplying the gaugino terms on the warped brane. 
Since already at the level of the effective perturbative superpotential 
we have found it consistent to promote $a$ to a chiral superfield,
$a \rightarrow e^{-T}$, then also here this part of the correction should be understood as a correction to the gauge kinetic function on
the warped brane, which now becomes
\beq \label{hwarped}
H_{warped}(S,T) = S + 2 b_0 T.
\eeq
The supergravity model defined with (\ref{wnpert}) and (\ref{hwarped})
gives effctive potential suitable to study moduli stabilization 
and supersymmetry breakdown due to gaugino condensation in the effective 
four-dimensional theory.\\
\vskip0.3cm 
The four-dimensional supergravity model defined by $K(S,\bar{S};T,\bar{T};\Phi,\bar{\Phi})=-M^{2}_P \log(S + \bar{S}) -3 M^{2}_P \log \left ( f(T+ \bar{T}
-\frac{k}{3 M^3} |\Phi_2|^2) -\frac{\beta k}{3 M^3} |\Phi_1|^2   \right )$, $W= 2 \sqrt{2} (W_1 + e^{-3 T} W_2 )$, 
$H_{warped}(S,T) = S + 2 b_0 T$ and $H_{planck}(S)=S$ 
has been constructed as a small perturbation around the generalized 
Randall-Sundrum background. However, it is very likely that extrapolating 
this model away from the original vacuum and using it on its own makes sense 
as a tool to explore phenomenology of warped compactifications at $TeV$ energies.  

Let us summarize basic features of our model. 
The F-terms take at the minimum the expectation values
\beqa &
|F^S|^2 = 8 e^K (S + \bar{S})^2 a^{2}(\pi \rho) |W_2|^2 (1-a^{2}(\pi \rho ))^2 \neq 0& \nn
&|F^T|^2 = 0& \nonumber \eeqa

which means that supersymmetry is broken along the dilaton direction.
The potential energy at this vacuum is negative:
\beq
V_{vac} = -\frac{8 |W_2|^2}{V_0 (M^3/k M^{2}_P)^3  M^{2}_P}
\frac{a^{2}(\pi \rho)}{1 - a^{2}(\pi \rho)}. 
\eeq
The mass of the canonically normalized radion is 
\beq
m^{2}_R = \frac{24}{V_0 (M^3/k M^{2}_P)^3 } \frac{a^{2}(\pi \rho) |W_2|^2}
{(1 -a^{2}(\pi \rho) )} \frac{1}{M^{4}_P}
\eeq
and the gravitino mass term is given by the expression 
\beq
m_{3/2} = \frac{2}{\sqrt{ V_0} (M^3/k M^{2}_P)^{3/2} } \frac{a(\pi \rho ) }{(1-a^{2}(\pi \rho ) )^{1/2}} \frac{|W_2|}{M^{2}_P}.
\eeq
It is interesting to compare these features to those of the no-scale models: 
there $F^S =0$, $V_{vac} =0$, and $F^T$ is undetermined at tree-level 
\cite{cfkn}.

\section{Summary}

The main result of this paper is the  four-dimensional 
effective supergravity action which describes low-energy physics 
of the Randall--Sundrum model with moduli fields 
in the bulk and charged chiral matter living on the branes.

The relation between 5d and 4d physics has been made explicit; the 
low-energy
action has been read off from a compactification of a locally supersymmetric model in five dimensions. 
The exponential warp factor has interesting consequences for the form of the 
effective 4d supergravity. 
The asymmetry between the warped and unwarped walls 
is visible in the K\"ahler function, gauge kinetic functions and in the
superpotential. Roughly speaking the contributions to these functions
which come from the warped wall are suppressed by an exponential factor
containing the radion superfield. This is the way the warp factor and (and RS brane tensions)
are encoded in the low-energy Lagrangian. 

We have described the mechanism of supersymmetry breaking mediation which relies on non-trivial configuration of the $Z_2$-odd  fields in the bulk. We point out
that the odd-$Z_2$-parity fields can be an important ingredient of 5d supersymmetric models. They play a
crucial role in communication between spatially separated branes. 

Moreover, we have demonstrated that, after freezing the dilaton, it is 
possible to stabilize the radion field in the backgrounds with broken
supersymmetry and excited odd-parity fields. To achieve this we do not
need to add vector fields and/or exotic charged matter in the bulk. 

We believe that the class of models we have constructed in this paper 
provides the proper explicit setup to study low-energy phenomenology of 
the supersymmetric brane models with warped vacua.    \\
\vskip0.5cm 
\noindent {\bf Acknowledgements}   
\noindent This work has been supported by RTN programs HPRN-CT-2000-00152
and HPRN-CT-2000-00148. 
Z.L. and S.P. are supported 
by the Polish Committee for Scientific Research under grant
5~P03B~119 20 (2001-2002).

\section*{Appendix A: The BPS solution in 5d supergravity coupled 
to the universal hypermultiplet}

First, let us  summarize the  basic facts about the geometry of the universal hypermultiplet.   
The 4 real coordinates are denoted $q^u=\{V, \sigma, x, y\}$; when convenient, we also use $\xi=x+iy$.
 
The metric is:
\bea
h_{VV}=\frac{1}{4V^2}& h_{\sigma\sigma}=\frac{1}{4V^2} & h_{\sigma V}=0
\nn
h_{xx}=\frac{1}{V}+\frac{y^2}{V^2} &  h_{yy}=\frac{1}{V}+\frac{x^2}{V^2} & h_{xy}=-\frac{xy}{V^2}
\nn
h_{\sigma x}=-\frac{y}{2V^2}& h_{\sigma y}=\frac{x}{2V^2}& 
\eea

The inverse metric:
\bea
h^{VV}=4V^2 &  h^{\sigma\sigma}=4V^2 + |\xi|^2  & h^{\sigma V}=0
\nn
h^{xx}=V &  h^{yy}=V+\frac{x^2}{V^2} & h^{xy}=0
\nn
h^{\sigma x}=2Vy& h_{\sigma y}=2Vx& 
\eea
 
We can introduce a vielbein one-form $V=V_u dq^u$. Metric can be expressed in terms of a vielbein as:
\beq
h_{uv}=\epsilon_{AB}\Omega_{ab}V_u^{Aa} V_v^{Bb}
\eeq
where $\epsilon$ and $\Omega$ are totally antisymmetric and we choose $\epsilon_{12}=\Omega_{21}=1$.
This formula determines the vielbein (up to $SU(2)$ transformation) to be:
\bea
V =\frac{1}{2\r2 V} \sigma^1 dV +\frac{1}{2\r2 V} \sigma^2 d\sigma +
 (\frac{1}{\sqrt{2V}}\sigma^3 +  \frac{y}{\r2 V}\sigma^2)dx + 
(\frac{1}{\sqrt{2V}}i I - \frac{x}{\r2 V}\sigma^2)dy
\eea
or more explicitly
\bea
V_V = \left (\begin{array}{cc}
    0                 &    \frac{1}{2\r2 V} \\
\frac{1}{2\r2 V}      &         0
            \end{array} \right )
&
V_\sigma = \left (\begin{array}{cc}
    0                 &    -i\frac{1}{2\r2 V} \\
i\frac{1}{2\r2 V}      &         0
            \end{array} \right )
\nn
V_x = \left (\begin{array}{cc}
\frac{1}{\sqrt{2V}}         &  -i\frac{y}{\r2 V} \\
 i\frac{y}{\r2 V}            &       -\frac{1}{\sqrt{2V}}   
            \end{array} \right )
&
V_y = \left (\begin{array}{cc}
i\frac{1}{\sqrt{2V}}         &  i\frac{x}{\r2 V} \\
-i\frac{x}{\r2 V}            &       i\frac{1}{\sqrt{2V}}   
            \end{array} \right )
\eea
The inverse vierbein $V^u$ satisfies the equation:
\beq
V^u_{Aa} V_v^{Aa}=\delta_v^u
\eeq
The explicit form is:
\bea
V^V = \left (\begin{array}{cc}
    0                 &    \r2 V \\
\r2 V      &         0
            \end{array} \right )
&
V^\sigma = \left (\begin{array}{cc}
     -i \sqrt{2V}\bar{\xi}           &    i\r2 V \\
-i\r2 V      &    -i \sqrt{2V}\xi     
            \end{array} \right )
\nn
V^x = \left (\begin{array}{cc}
 \sqrt{\frac{V}{2}}          &   0\\
 0      &       -\sqrt{\frac{V}{2}}   
            \end{array} \right )
&
V^y = \left (\begin{array}{cc}
   -i\sqrt{\frac{V}{2}}         &  0\\
0           &       -i\sqrt{\frac{V}{2}}   
            \end{array} \right )
\eea

The SU(2) spin connection $\omega$ : 
\bea
\omega_V=0  
&
\omega_\sigma=\frac{1}{4V} i \sigma^3
\nn
\omega_x= \frac{y}{2V} i\sigma^3 - \frac{1}{V^{1/2}}i\sigma^2
&
\omega_y= -\frac{x}{2V} i\sigma^3 - \frac{1}{V^{1/2}}i\sigma^1
\eea
and the Sp(1) spin connection $\Delta$:
\bea
\Delta_V=0
&
\Delta_\sigma=-\frac{3}{4V} i \sigma^3
\nn
\Delta_x= -\frac{3y}{2V} i\sigma^3
&
\Delta_y= \frac{3x}{2V} i\sigma^3
\eea

The K\"ahler form (which for quaternionic manifolds that can occur in supersymmetric theories is minus the curvature form) is: 
\bea
K_{V\sigma}=\frac{1}{8V^2}i\sigma^3
&
K_{Vx}=-\frac{1}{4V^{3/2}}i\sigma^2 + \frac{y}{4V^2}i\sigma^3
&
K_{Vy}=-\frac{1}{4V^{3/2}}i\sigma^1 - \frac{x}{4V^2}i\sigma^3
\nn
K_{\sigma x}=\frac{1}{4V^{3/2}}i \sigma^1
&
K_{\sigma y}=-\frac{1}{4V^{3/2}}i \sigma^2
&
K_{xy}=-\frac{1}{2V} i \sigma^3 +  \frac{x}{2V^{3/2}}i\sigma^1 -\frac{y}{2V^{3/2}}i\sigma^2
\eea

To illustrate our discussion of supersymmetry breaking let us study the explicit model presented in \cite{flp2}. It is based on  5d dimensional supergravity with the universal hypermultiplet and the U(1) symmetry $\xi \ra e^{i\phi} \xi $ gauged. The Killing vector and prepotential corresponding to our gauging have  components: 
\bea
k^x=-y \; k^y=x \; k^V=k^\sigma=0
\nn
P_1=-\frac{x}{2\sqrt{V}}\; P_2=\frac{y}{2\sqrt{V}} \; P_3=\frac{1}{4}(1-\frac{|\xi|^2}{V})
\nn
\sqrt{\vec{P}^2} =\frac{1}{4}(1+\frac{|\xi|^2}{V}) 
\eea  
Locally, even with  non-zero sources for the odd field $\xi$ it is possible to find a BPS solution, which preserves half of the supersymmetry. 
Plugging in the explicit metric and the prepotential into the BPS conditions (\ref{hhBPS}) we get:
\bea
(V)&
 V' = -6k R_0 |\xi|^2&
\nn 
(\sigma)&
\sigma ' -2 (yx'-xy') =0
&\nn 
(x)&
x' - \frac{y}{V}(\sigma ' -2 (yx'-xy'))= 3k R_0 x
&\nn
(y)&
y' +\frac{x}{V}(\sigma ' -2 (yx'-xy'))= 3k R_0 y
&\nn&
\frac{a'}{a}=-k R_0 (1+\frac{|\xi|^2}{V})
&  
\eea   
It is straightforward to find the solution:
\bea
x=C_1 \exp (3kR_0|x^5|) \epsilon(x^5)
\nn
y=C_2 \exp (3kR_0|x^5|) \epsilon(x^5)
\nn
V= V_0 - (C_1^2+C_2^2) \exp (6k R_0|x^5|) 
\nn
\sigma=\sigma_0
\eea       
Integrating equation for the warp factor we  find:
\beq
a(x^5)=\exp (-kR_0|x^5|)(V_0 - (C_1^2+C_2^2) e^{3 k  R_0 |x^5|})^{1/6}
\eeq
The matching condition in the Einstein equation yields:
\bea
\label{to}
\frac{a'}{a}(0)=-kR_0 (1-\frac{|\xi|^2}{V}(0)), &  
\frac{a'}{a}(\pi\rho)=-k(1-\frac{|\xi|^2}{V}(\pi\rho))
\eea
If $C_1=C_2=0$, the above set is satisfied and the flat BPS solution satisfies all boundary conditions. But as soon as we induce a non-zero value of $\xi$ (by coupling $\pa_5 \xi$ to the boundary sources) it is impossible to find any solution to (\ref{to}) and supersymmetry is broken.

\section*{Appendix B: Supersymmetric coupling of\\
the odd hypermultiplet fields
to the brane sources} 

In this appendix we investigate the supersymmetric coupling  of the hypermultiplet odd fields to the brane sources.
It is obvious that away from the branes the theory does not distinguish between the even and odd fields. However, matching conditions at the fixed points change in a significant way. To see this we factorize the $x^5$ dependence of an odd field $\xi$:
\beq
\xi(x^5)= \epsilon(x^5)\zeta(|x^5|),
\eeq
where $\zeta$ is smooth near the fixed points; hence:
\bea
\pa_5 \xi = \zeta' + 
2 (\delta(x^5)\zeta(0)-\delta(x^5-\pi \rho) \zeta(\pi \rho))
\nn
\pa_5\pa_5 \xi = \zeta''\epsilon(x^5) + 2 ( \delta '(x^5)\zeta(0)
-\delta'(x^5-\pi \rho) \zeta(\pi \rho)).
\eea
The odd fields that are non-zero at the $Z_2$ fixed points must have a jump there. As a consequence their first derivatives have delta functions singularities and their second derivatives  have derivatives of the delta. 
The equations of motion are second order, thus we must somehow cancel 
$\delta'$ singularities by a suitable choice of the brane action.
One way to achieve this is to couple the odd fields to the sources on the boundary through their fifth derivative. Such couplings naturally arise in M-theory \cite{hw} and its compactification to 5d \cite{ovrut} where the field $\xi$ from the universal hypermultiplet couples to gaugino bilinears and to the superpotential of the boundary scalars. 

Local supersymmetry considerably restricts  the possible form of such derivative couplings to the brane sources. We will show that couplings of the fifth derivative of an odd field to the boundary sources $\chi^r$ has to enter the bosonic action (\ref{hyperaction}) in a form of 'the full square'. That is we have to replace $\pa_5 q^r \rightarrow \hat{\pa}_5 q^r \equiv \pa_5 q^r + \delta(x^5)(\chi^r)_1 + \delta(x^5-\pi\rho)(\chi^r)_2$, in the kinetic terms of the odd fields (note that it implies that $\delta^2$ singularities appear in the lagrangian).     

Let us introduce the following coupling of an odd scalar $q^r$ to the source $\chi$ located on, say, the hidden brane:
\beq
\label{lksi}
\cl_\xi=\delta(x^5)\frac{e_4}{e_5^5}\pa_5 q^r \chi_r (q_u)      
\eeq
We allow for the dependence of  the source $\chi$ on the even hypermultiplet scalar (it will turn out crucial for the consistency). Now we want to supersymmetrize (\ref{lksi}).
Consider the variation of the (odd) scalar $q$ in $\cl_\xi$:
\bea
\delta q^r= \frac{i}{2} V^r_{Aa}\ov{\epsilon}^A \lambda^a
\nn
\ra \delta \cl_\xi =\frac{e_4}{e_5^5}  
\frac{i}{2} \pa_5 (V^r_{Aa}\ov{\epsilon}^A \lambda^a) \chi_r 
\eea
First consider the fifth derivative acting on $\epsilon$. Using the fact that $\delta \psi_5^A =\pa_5 \epsilon^A$ this variation can be cancelled by adding a new term to the boundary lagrangian:
\beq
\label{l5}
\cl_5 = -\delta(x^5)\frac{e_4}{e_5^5}  
\frac{i}{2} V^r_{Aa}(\ov{\lambda}^a\psi_5^A) \chi_r 
\eeq 
The variation with the 5th derivative acting on $\lambda$ can be cancelled by the variation of the hyperino kinetic term provided we modify the hyperino transformation law in the following way:
\bea
\delta \lambda^a = -\frac{\delta(x^5)}{e_5^5}\frac{i}{2}\Omega^{ab} V^s_{bB}\gamma^5\epsilon^B \chi_s
\eea
This modification has an immediate consequence that $\delta^2$ terms are necessary in the brane lagrangian because $\lambda$ is already present in the brane lagrangian. Namely, varying (\ref{l5}) we get:
\bea
\delta  \cl_5 =
 (\delta(x^5))^2\frac{e_4}{(e_5^5)^2}\frac{1}{4} V^r_{Aa} \Omega^{ab} V^s_{bB} \ov{\psi_5}^A\epsilon^B \chi_r\chi_s=
\nn
(\delta(x^5))^2\frac{e_4}{(e_5^5)^2}\frac{1}{8}(\ov{\psi_5}^A \gamma^5\epsilon_A) h^{rs} \chi_r\chi_s 
\eea
Recalling, that $\delta e_5^5 = -\frac{1}{2}(\ov{\psi_5}^A \gamma^5\epsilon_A)$ the above can be cancelled by adding a singular term to the brane lagrangian:
\beq
\label{ld2}
\cl_\delta = -(\delta(x^5))^2\frac{e_4}{e_5^5}\frac{1}{4} h^{rs}\chi_r\chi_s   
\eeq
This is exactly what we need to cancel the $\delta^2$ singularities in the equations of motion.

Note that (\ref{lksi}), (\ref{ld2}) and the scalar kinetic term can be gathered in a full square:
\beq
\cl_{FS}=e_5 g^{\alpha \beta} h_{uw} \hat{\pa_\alpha q^u} \hat {\pa_\beta q^w}
\eeq
where we defined :
\beq
\hat{\pa_\alpha q^u}= \pa_\alpha q^u -\delta_\alpha^5 \frac{1}{2} h^{ur}\chi_r \delta (x^5)
\eeq
   
Thus we can {\it guess} that all the terms needed to supersymmetrize the action can be found by exchanging in the bulk lagrangian the fifth derivative of the scalars by the hatted one. This procedure leads to appearance of the following new terms in the brane lagrangian:
\bea
\cl_B = e_4 \delta (x^5) \left ( 
-\frac{i}{2} \ov{\lambda}^a\gamma^\alpha \gamma^5 \psi_\alpha^A V^r_{Aa} 
+\frac{1}{4} \ov{\psi_\mu}_A\gamma^{\mu\nu}\gamma^5 \psi_\nu^B (\omega_r)^A_B h^{rs}\chi_s
- \frac{1}{4}\ov{\lambda}^a\gamma^5\lambda^b(\Delta_r)^a_b h^{rs} \chi_s 
\right)       
\eea
and in the SUSY transformation laws:
\bea
&
\delta \psi_5^A = -\frac{1}{2}(\omega_r)^A_B\epsilon^B h^{rs}\chi_s \delta(x^5) 
&\nn&
 \delta \lambda^a = -\frac{i}{2}\gamma^5\epsilon^B \Omega^{ab} V^s_{bB}\chi_s\frac{\delta(x^5)}{e_5^5}
&
\eea
Note that this procedure correctly reproduces the terms we determined before, 
e.g. the correction to the hyperino transformation laws. 

Nevertheless, exchanging the derivatives with their hatted counterpart does not ensure that the action is supersymmetric. If we make the supersymmetry 
transformation of  the modified lagrangian we get:
\bea
&\delta \cl (\phi, \hat{\pa_5 q^u}, \hat{\pa_5 q^u}\hat{\pa_5 q^w}) =   
\frac{\pa \cl}{\pa \phi}\delta\phi+
\frac{\pa \cl}{\pa \hat{\pa_5 q^u}}\delta\hat{\pa_5 q^u}+ 
2\frac{\pa \cl}{\pa \hat{\pa_5 q^u}\pa\hat{\pa_5 q^w}}\delta\hat{\pa_5 q^u}\hat{\pa_5 q^w}
&\nn&
\frac{\pa \cl}{\pa \phi}\delta\phi+\frac{\pa \cl}{\pa \hat{\pa_5 q^u}}\delta{\pa_5 q^u}+2\frac{\pa \cl}{\pa \hat{\pa_5 q^u}\pa\hat{\pa_5 q^w}}\delta{\pa_5 q^u}\hat{\pa_5 q^w}
-\frac{1}{2}\delta(h^{ur}\chi_r) \delta (x^5) (
\frac{\pa \cl}{\pa \hat{\pa_5 q^u}}+ 2\frac{\pa \cl}{\pa \hat{\pa_5 q^u}\pa \hat{\pa_5 q^w}}\hat{\pa_5 q^w}).
&\eea

The first three terms vanish here if they vanish in the unmodified 5d supergravity. But the last term does not vanish automatically. The easiest way to nullify this contribution is to assume that the supersymmetry variation $\delta(h^{ur}\chi_r)$ is identically zero. This can be achieved if the function $\chi$ which determines the coupling of odd scalars to the boundary can be expressed as:
\beq
\label{chi}
\chi_r= h_{rs} A^s
\eeq
where $A^s$ are constants (which determine  the boundary values of the odd fields). 

Let us turn to  the example of the universal hypermultiplet. The equation (\ref{chi}) implies that the coupling of the odd scalars to the brane has the form:
\beq
\cl_\xi = \frac{A^x}{V}(1+\frac{y^2-xy\frac{A^y}{A^x}}{V}) \pa_5 x +
 \frac{A^y}{V}(1+\frac{x^2-xy\frac{A^x}{A^y}}{V}) \pa_5 y .
\eeq

\section*{Appendix C: Supersymmetric coupling of 
gauge sectors on the boundaries to the bulk} 

For completeness let us summarize  the relevant parts of the 
bulk bosonic action
and the brane action coupled to it. Derivation of the complete result 
including four-fermions terms can be found in \cite{mt}. 
The action is $S = S_{bulk} + S_{YM} + S_{matter}$ where
\beqa \label{eq:acc}
&S_{bulk} = \int d^5 x \; e_5 (\frac{R}{2} 
-\frac{1}{4V^2}(\pa_{\alpha}V\pa^{\alpha}V + \pa_{\alpha}\sigma \pa^{\alpha}
\sigma -\frac {1}{V}\pa_{\alpha}\xi\pa^{\alpha}\bar{\xi} 
+ 6 k^2 (1 + \frac{1}{2 V} 
|\xi|^2 - \frac{1}{2 V^{2} } |\xi|^4 ))& \nonumber \\
&S_{YM\;i}= \int d^5x \frac{e_5}{e_5^5}\delta(x^5-x^{5}_i) 
\left ( \right .  
 -\frac {V} {4} 
F_{\mu\nu}^{a} F^{a\mu\nu}   
-\frac{1}{4} 
\sigma F_{\mu\nu}^{a} \tilde F^{a\mu\nu} 
-\frac {V} {2} 
\overline{\chi^{a}}D\!\!\!\!\slash\chi^{a} 
+\frac{V}{4}  
(\overline{\psi}_{\mu} \gamma^{\nu\rho}\gamma^{\mu}\chi^{a})F^{a}_{\nu\rho} & \nonumber \\
&+\frac {3i} {4\sqrt{2}}\frac {V}{e_{5}^{5}} 
(\overline{\chi}^{a}\gamma^{5}\gamma^{\mu}\chi^{a}){\cal F}_{\mu 5}
-\frac {1} {4} 
(\overline{\lambda}\gamma^{\nu\rho}\chi^{a})F_{\nu\rho}^{a} 
-\frac {i} {8} 
(\overline{\chi}^{a}\gamma^{5}\gamma^{\mu}\chi^{a})\partial_{\mu}\sigma 
& \nonumber \\
&-\frac{\sqrt{V}}{2e_{5}^{5}} 
( 
(\overline{\chi^{a}}_{L}\chi^{a}_{R})\partial_{5}\overline{\xi} 
+ (\overline{\chi^{a}}_{R}\chi^{a}_{L})\partial_{5}\xi 
) 
+ \delta(0) 
\frac{V^{3/2}}{8} ((\chi^{a})^2)^2 )
+  ({\rmfamily 4-fermions}) 
\left .   \right )& \nonumber \\
&S_{matter\; i}= \int d^5 x \; \frac{e_5}{e^{5}_{5}}  \delta(x^5- x_i ) ( -
\epsilon_i 6 k 
(1-\frac{|\xi|^2}{V})  -  \frac{2}{V} \sqrt{g^{55}}
(W \xi' + \bar{W}_i \bar{\xi'}  & \nonumber \\
&-D_\mu \Phi_i D^\mu
\bar{\Phi}^{i} - \frac{4}{V} \frac{\partial W_i}{\partial \Phi_i} 
\frac{\partial \bar{W}_i}{\partial \bar{\Phi}^{i}}
+ \delta(0) (4 W \bar{W} + V^{3/2} \bar{W} (\ov{\chi^{a}_R} \chi^{a}_L))  
&\nn&
-(\frac{W}{\sqrt{V}}
(\ov{\psi}_{L \mu} \gamma^{\mu \nu} \psi_{R\nu}) - \frac{1}{V^{2/3}} W ( 
\ov{\psi}_{L \mu} \gamma^\mu \lambda_L ) + \frac{i}{V^{3/2}} e^{5}_5 W
(\ov{\psi}_{R5} \lambda_L) + h.c.)
 ) &
\eeqa
where $i=1,2$ labels branes and $\epsilon_{1,2}=+1,-1$. 
In the above $Z_2$-even 4d Majorana fermions are defined as:  
\bea 
\label{spinorss}  
 \psi_{\mu} = \left ( 
\begin{array}{cccc} 
{\psi_{L\mu}^{2}} \\ 
 {\psi_{R\mu}^{1}} 
\end{array} 
\right ), 
& 
 \lambda =\sqrt{2}V \left ( 
\begin{array}{cccc} 
{-i\lambda_{L}^{1}} \\ 
 {i\lambda_{R}^{2}} 
\end{array} 
\right ),  
\eea   
and brane supersymmetry is generated by the Majorana fermion 
$\epsilon = \left ( 
\begin{array}{cccc} 
{\epsilon_{L}^{2}} \\ 
 {\epsilon_{R}^{1}} 
\end{array} 
\right ) $. 
The relevant parts of the supersymmetry transformation of bulk fermions,
which depend explicitly on brane operators that are allowed to take 
an expectation value, are
\beqa
&\delta \psi_5 = \frac{2 i e^{5}_5}{\sqrt{V}} (W_i \epsilon_L - \bar{W}_i  
\epsilon_R)
 \delta(x^5 - x^{5}_i)& \nonumber \\
& \delta \lambda = -2 \sqrt{V} (W_i \epsilon_L - \bar{W}_i  \epsilon_R)
 \delta(x^5 - x^{5}_i).& 
\eeqa
Supersymmetry transformations of the charged fermions on branes are 
$\delta \chi_\Phi = \frac{1}{2} ( D_\rho \Phi^p - \bar{\psi}_{R \rho} \chi_\Phi^p) \gamma^\rho \epsilon_R + \frac{1}{8 V} \chi_{\Phi\,L} (\bar{\epsilon} \gamma^5 \lambda) - \frac{1}{\sqrt{V}} \frac{\partial \bar{W} }{\partial \bar{\Phi}_p} \epsilon_L$.  
After solving the equations of motion for $\xi$, boundary operators appear also in bulk parts of the modulini supersymmetry transformations (these are 
given in the main text).  \\
\section*{Appendix D : Conventions and normalizations}

The normalizations are mainly those of \cite{ovrut}.
The signature of the metric tensor is $(-++++)$. The indices $\alpha, \beta ...$  are  five-dimensional ($0,...,3, 5$), while 4d indices are denoted by $\mu, \nu, ...$. We define $\gamma^5 = \left (\begin{array}{cc} -1 & 0 \\ 0 & 1  \end{array} \right )$.

  The  $Z_2$ symmetry acts as  reflection $x^5 \rightarrow -x^5$ and is represented in such a way  that bosonic fields $(e_{\mu}^{m},e_{5}^{5}, {\cal A}_{5})$ are even, and $(e_{5}^{m},e_{\mu}^{5}, {\cal A}_{\mu})$ are odd. The action of $Z_2$ on the gravitino is 
$\gamma^5 \psi_\mu^A(x^5) = (\sigma^3)^A_{\;B}\psi_\mu^B(-x^5)$
$\gamma^5 \psi_5^A(x^5) = -(\sigma^3)^5_{\;B}\psi_5^B(-x^5)$ (so that, for instance, $\psi_{\mu R}^1$ is even and $\psi_{5 R}^1$ is odd). The action of $Z_2$ on the SUSY parameter $\epsilon^A$ is the same as the action on $\psi_\mu^A$.

In the spinor basis we use, the SU(2) R-symmetry is manifest. The index $A$ of the gravitino transforms in the fundamental representation of SU(2). 
The SU(2)  indices are raised with an antisymmetric tensor $\epsilon^{AB}$.
We choose $\epsilon^{12}=\epsilon_{12}=1$. Explicitly: $\psi^1 = \psi_2$ $\psi^2 = -\psi_1$.  Note that $\ov{\psi}^A \equiv \ov{\psi_A}$ and $\ov{\psi}_A \equiv -\ov{\psi^A}$, so one has to be careful about the position of the bar.
 The Majorana condition is:
\beq
\ov{\psi_\alpha}^A = (\psi_\alpha^A)^T C,
\eeq
where $C$ is the 5d charge conjugation matrix satisfying $C\gamma^\alpha C^{-1} = (\gamma^\alpha)^T$. In particular $\psi_\alpha^1 = -C_5 (\ov{\psi_\alpha^2})^T$, $\psi_\alpha^2 = C_5 (\ov{\psi_\alpha^1})^T$. In the chiral basis $C=i \gamma^2\gamma^0 \gamma^5$.   

The rule for dealing with symplectic spinors is $\ov{\psi}^A\gamma^{\mu_1 \dots \mu_n}\chi^B=\ov{\chi}^B\gamma^{\mu_n \dots \mu_1}\psi^A$. From the above formula one can deduce:  $\ov{\psi_\mu^1} \gamma ^\mu \epsilon^1 = \ov{\psi_{\mu 2}} \gamma ^\mu \epsilon^1 =\ov{\psi_\mu}^2 \gamma ^\mu \epsilon^1=\ov{\epsilon}^1 \gamma ^\mu \psi_\mu^2= \ov{\epsilon_1} \gamma ^\mu \psi_\mu^2 = -\ov{\epsilon^2} \gamma ^\mu \psi_\mu^2$.\\

\end{document}